\documentclass[11pt]{article}

\usepackage{amsmath}
\usepackage{epsfig}
\usepackage{graphics}

\newcommand{\N}{N\raise.7ex\hbox{\underline{$\circ $}}$\;$}

\begin{document}

\title{
 V.M. Red'kov\footnote{redkov@dragon.bas-net.by}, E.A. Tolkachev \footnote{tea@dragon.bas-net.by} \\
Quaternions and Small  Lorentz Groups \\ in Noncommutative Electrodynamics \\[5mm]
{\small B.I. Stepanov Institute of Physics \\
National Academy of Sciences of Belarus}  }

\maketitle

\begin{quotation}

Non-linear electrodynamics arising in the frames of  field
theories in noncommuta\-ti\-ve space-time is examined on the base
of quaternion
  formalism.
The problem of form-invariance of the corresponding nonlinear
constitutive relations governed by six noncommutativity
parameters $\theta_{kl}$ or quaternion $  \underline{K}
 = \underline{\theta} - i \underline{\epsilon} $ is explored in detail.
Two  Abelian 2-parametric small groups,  $SO(2) \otimes SO(1.1)$
or $T_{2}$,  depending on
  invariant length   $ \underline{K}^{2}\neq 0$ or $ \underline{K}^{2}= 0$ respectively, have been found.
The way to interpret both   small groups in physical terms
consists in factorizing  corresponding Lorentz transformations
into Euclidean rotations and  Lorentzian boosts.

In the context  of general  study of various dual symmetries  in
noncommutative field theory,
  it is demonstrated explicitly  that
the non-linear constitutive equations under consideration are not
invariant under continuous dual rotations, instead only invariance
under  discrete dual transformation exists.

\end{quotation}

\section{Introduction}

It was James Clerk Maxwell  who first used the quaternionic
algebra by  Sir William Rowan Hamilton to deal with
electrodynamic equations. Since then, the  areas of applying of
quaternions steadily extend. Bi-quaternion algebra, that is  the
algebra by William Kingdon Clifford in 2-dimension complex space,
 turns to be especially useful in physical applications.
The ground for so extensive employing the quaternions is that this
formalism provides us with simple algebraic  tools to deal with
spinors of relativistic physics without any use of the  cumbersome
index  technique \cite{Berezin-Kurochkin-Tolkachev-1989}.

The article aims to demonstrate the effectiveness of this
algebraic technique to approach symmetries of Maxwell-like
equations, equivalent in the sense by  Seiberg -- Witten
\cite{Seiberg-Witten-1999} to the electrodynamic equations in  a
noncommutative space-time. As known \cite{Douglas-Nekrasov-2001},
\cite{Seiberg-Witten-1999},
\cite{Chaichian-Kulish-Nishijima-Tureanu-2004},
 interest  in field theory models in a noncommutative  space-time
has been grown notably after creating in
\cite{Seiberg-Witten-1999}  a general algorithm to relate  usual
Yang-Mills gauge models to their noncommutative  counterparts.
There appears  a great deal of new physical problems to
investigate, besides the question of the hypothetic  coordinate
non-commutativity has become   of  practically  testable nature.
Noticeable progress in describing  symmetry of noncommutative
spaces  was achieved  on the base of twisted Poincar\'e group
\cite{Chaichian-Kulish-Nishijima-Tureanu-2004},
\cite{Alvarez-Vazquez-2003}, \cite{Fiore-Wess-2007},
\cite{Chaichian-Kulish-Nishijima-Tureanu-2008}.

For instance, the mapping by Seiberg -- Witten refers  the
noncommutative extension of electro\-dynamics  to the usual
microscopic Maxwell  theory with  special nonlinear constitutive
relations. Examining all possible sym\-metries of these new
constitutive relations seems to be a  significant point in order
to discern  the effects of the space-time  non-commutativity in
observable electromagnetic    non-linear  effects.

The problem of form-invariance of the non-commutativity structural
equations (see below) was considered in
\cite{Alvarez-Vazquez-2003},\cite{Chaichian-Kulish-Nishijima-Tureanu-2008}.
Several simple noncommutative parameters were listed which alow
for existence of some residual Lorentz symmetry -- the later is
recognized  to have the structure $SO(2) \otimes SO(1.1)$. In was
claimed in \cite{Alvarez-Vazquez-2003} that in the case of an
arbitrary noncommutative matrix no residual Lorentz symmetry
exists.

As known, the commutator  for space-time coordinates $\hat{x}^{a}$ in the Weyl-Moyal space
is defined by an antisymmetric matrix $\theta^{ab}$ elements of which are real numbers:
\begin{eqnarray}
[ \hat{x}^{\mu}, \hat{x}^{\nu} ] = i \; \theta^{\mu \nu}
\;.
\nonumber
\end{eqnarray}

To any operator $f(\hat{x})$  there corresponds the Weyl symbol $f(x)$ defined on commutative  Minkowski
space with coordinates $x^{a}$. To the product of  operators corresponds an operation $*$
\begin{eqnarray}
(f * g)(x) =
\left.
 f(x) \; \mbox{exp} \; ({i \over 2}  \theta^{\mu \nu} \overleftarrow{\partial}_{\mu}
 \overrightarrow{\partial}_{\nu} )\; g(x') \right |_{x'=x}\; .
\label{1}
\end{eqnarray}

\noindent From (\ref{1}) it follows

\begin{eqnarray}
\; [ x^{\mu} , x^{\nu} ] _{*} = x^{\mu} * x^{\nu} - x^{\nu} * x^{\mu} =  i \; \theta ^{\mu \nu} \; .
\label{2}
\end{eqnarray}

Consistent use  of the operation $*$ permits to define twisted analogue  for Poincar\'e algebra [3-6]
and provides us with possibility to construct representation of the Poincar\'e group.
It was shown in \cite{Chaichian-Kulish-Nishijima-Tureanu-2008} that at special
 particular choice of $\theta^{\mu \nu }$ " ... the twisted Poincar\'e symmetry
 of noncommutative (quantum) field theory is reduced to the residual $O(1,1) \otimes S0(2)$ symmetry, but
still carrying representations of the   full Lorentz group.  ... The meaning of the twisted  Poincar\'e symmetry
in NC QFT becomes transparent:
it represents actually the invariance with respect to the stability group of $\theta_{\mu \nu }$,
while  the quantum  fields carry representations of the full Lorentz group and the  Hilbert space of states has the richness of
particle representations of the commutative QFT".

In this context it is important to have known what is the full residual  Lorentz symmetry for
arbitrary matrix $\theta_{ab}$; because it was claimed in \cite{Alvarez-Vazquez-2003}
that in the case of arbitrary noncomutative matrix no residual symmetry exists.
There exist several different views  \cite{Amelino...-2007} on the transforms of the matrix $\theta^{\mu\nu}$
under the Lorentz group, most radical attitude is to consider $\theta_{ab}$ as new fundamental constants like
the minimal length. In fact, our consideration  does not depend on these different views.

In the paper, we use an  quaternionic technique as a main tool for
describing  all  Lorentz subgroups leaving invariant any 2-rank
antisymmetric tensor. The treatment is given in  the most general
form irrespective of  explicit  vector constituents of the
tensor. We place these  quite  conventional mathematical facts in
the context of    Maxwell equations in noncommutative
space-time. As known according to of Seiberg -- Witten
\cite{Seiberg-Witten-1999},   in the first order approximation in
parameters $\theta^{\mu\nu}$ we get the ordinary Maxwell equations
and special nonlinear constitutive relations. Just the symmetry
properties of these nonlinear constitutive relations is the main
subject of the paper. It permits us to to make more clear and full
the known results on residual Lorentz symmetry  $SO(2) \otimes
SO(1.1)$.

As mentioned, several particular examples of such small  (or
stability)  subgroups    were noticed  in the literature, so our
analysis extends and completes  previous considerations. In a
sense, the problem  may be   solved  with the
help of old and well elaborated technique in the theory of the
Lorentz group.
In the main parts, our examination of the problem
is based on  the use of bi-quaternion formalism
\cite{Berezin-Kurochkin-Tolkachev-1989}. Brief translation to the
more traditional technique \cite{Fedorov-1980} developed for the
Lorentz group and related to it will be given too, when instead of
the antisymmetric tensor $\theta_{\mu\nu}$  we use  a
corresponding
 3-vector under the complex
orthogonal group SO(3.C) which is equivalent to a symmetrical
2-rank spinor
 under SL(2.C)\footnote{More detailed treatment
of the problem in the frame of
Rieman-Zilberstein-Majorana-Oppenheimer  formalism and
conventional spinor formalism
 will be published  elsewhere.}.

In the context  of general  study of various dual symmetries  in
noncommutative field theory \cite{Aschiery-2001},
\cite{Ganor-Rajesh-Sethi-2000}, \cite{Reya-Ungeb-2007}, one other
problem will be considered:  it is demonstrated explicitly  that
the known nonlinear constitutive equations arising from
noncommutative  electrodynamics in  the first order approximation
are not invariant under continuous dual rotations, instead only
invariance under  discrete dual transformation exists, which
contrasts with claim of the paper \cite{Aschiery-2001}.

\section{ Residual Lorentz symmetry of the noncommutative  Maxwell theory,   quaternion treatment  }

It is known that extended Maxwell equations in noncommutative
space-time, by means of Seiberg -- Witten map in the first order
approximation in parameters $\theta^{\mu\nu}$ provide us with the
ordinary  Maxwell equations and special nonlinear  constitutive
relations.
\begin{eqnarray}
\mbox{div} \; \vec{ B} = 0 \; , \qquad \mbox{rot} \;\vec{E} =
-{\partial \vec{B} \over \partial   t} \; ; \label{1.2a}
\\
\mbox{div}\; \vec{D} = 0\; , \qquad
 \mbox{rot} \;  \vec{H}  =    {\partial \vec{D} \over \partial  t} \; ,
\label{1.2b}
\end{eqnarray}

\noindent and constitutive relations
\begin{eqnarray}
\vec{D}  = \vec{E} + [ \; ( \vec{\epsilon} \vec{E}) -  (
\vec{\theta} \vec{B})\;  ]\; \vec{E} +
 [ \; ( \vec{\theta} \vec{E}) +  (\vec{\epsilon} \vec{B})\;  ]\; \vec{B} +
 (\vec{E} \vec{B})\;
  \vec{\theta}  + {1\over 2}(\vec{E}^{2}-
 \vec{B}^{2}) \; \vec{\epsilon} \; ,
\nonumber
\\
 \vec{H} =  \vec{B} + [ \; ( \vec{\epsilon} \vec{E}) -  ( \vec{\theta} \vec{B})\;  ]\; \vec{B} -
 [ \; ( \vec{\theta} \vec{E}) +  ( \vec{\epsilon} \vec{B})\;  ]\; \vec{E} -
 ( \vec{E} \vec{B})\;   \vec{\epsilon}   + {1\over 2}( \vec{E}^{2}-
 \vec{B}^{2}) \; \vec{\theta} \; ,
\label{1.3a}
\end{eqnarray}

\noindent and inverse relations (within the accuracy of first
order terms)
\begin{eqnarray}
 \vec{E} = \vec{D}  +
  [ \; \vec{\theta} \vec{H}  - \vec{\epsilon} \vec{D} \; ] \; \vec{D} -
[ \; \vec{\theta} \vec{D} + \vec{\epsilon} \vec{H}  \;  ] \;
\vec{H} - ( \vec{D}  \vec{H} ) \; \vec{\theta}  +{1 \over 2} \; (
\vec{H}^{2}  - \vec{D}^{2} ) \; \vec{\epsilon} \; , \nonumber
\\
 \vec{B} = \vec{H}  +
  [ \; \vec{\theta} \vec{H}  - \vec{\epsilon}  \vec{D}  \; ] \; \vec{H} +
[ \; \vec{\theta} \vec{D} + \vec{\epsilon} \vec{H} \;  ] \;
\vec{D}  +
 ( \vec{D} \vec{H} \; \vec{\epsilon}  +{1 \over 2}
\; ( \vec{H}^{2} - \vec{D}^{2}  ) \; \vec{\theta} \; .
\label{1.3b}
\end{eqnarray}

\noindent The conventional notation is used:
\begin{eqnarray}
E^{m} = F^{m0}\; , \;\;  B^{k} = -{1 \over 2} \epsilon^{klm}
F_{lm}\; , \qquad \epsilon^{m} = \theta^{m0}\; , \;\;  \theta^{k}
= -{1 \over 2} \epsilon^{klm} \theta_{lm}\; ; \nonumber
\end{eqnarray}

\noindent  $\vec{\epsilon}$ and $\vec{\theta}$ appear to be
parameters of effective nonlinear media. In general, each of
equations (\ref{1.2a}) and (\ref{1.2b}) exhibits a 20-dimensional
Lie group  symmetry \cite{Nikitin-1983}; in which manner the
presence of the nonlinear constitutive equations (\ref{1.3a}) and
(\ref{1.3b})  must constrict
 this symmetry. This is a main question.
  In solving the problem we will apply quaternion  technique \cite{Berezin-Kurochkin-Tolkachev-1989}.

Any element in bi-quaternion algebra  (algebra over complex
numbers) can be presented as\footnote{Ordinary 3-vectors are
referred as $\vec{a}$, whereas $\underline{p}$ designates  a
vector part of a quaternion.}
\begin{eqnarray}
q = q_{0} e_{0} +  q_{a} \; e_{a} = q_{0} + \underline{q} =
q_{s} + q_{\mbox{v}}\; , \nonumber
\end{eqnarray}

\noindent where basic quaternions obey
\begin{eqnarray}
e_{0} e_{a} = e_{a} e_{0}\; , \qquad  e_{0}^{2} = e_{0}, \qquad
e_{a} e_{b} = - \delta_{ab} \; e_{0}+ \epsilon_{abc} \; e_{c}\; , \nonumber
\end{eqnarray}

\noindent so the product of two quaternions is given by
\begin{eqnarray}
q\; p = (q_{0} \; p_{0} - \vec{q} \; \vec{p} )  \; e_{0}+ (q_{0}\;
\vec{p} +  p_{0} \; \vec{q}  + \vec{q} \times \vec{p}) \;
\vec{e}\; . \nonumber
\end{eqnarray}

\noindent Two special operations for bi-quaternions are defined:
quaternion conjugation
\begin{eqnarray}
\bar{q} =  q_{0} - e_{a}\; q_{a} = q_{0} - \underline{q}\; ,
\qquad   q\; \bar{q} = \bar{q}\; q = q_{0}^{2} + \vec{q}^{\;2} \;,
\qquad  \stackrel{-}{(q p)}= \bar{p} \;  \bar{q}  \nonumber
\end{eqnarray}

\noindent and complex conjugation
\begin{eqnarray}
q^{*} = q_{0}^{*} - q_{a}^{\;*} e_{a} =  q_{0}^{*} -
\underline{q}^{*} \; . \nonumber
\end{eqnarray}

\noindent With the use of notation
\begin{eqnarray}
\nabla = -i \partial_{t} e_{0}  +  \underline{e} \; \nabla =   -i \partial_{t}
+ \underline{\nabla}\; , \nonumber
\\
\underline{f} = \underline{B} - i \underline{E} \; ,\qquad
\underline{h} = \underline{H} - i \underline{D}
\end{eqnarray}

\noindent Maxwell equations in media can be presented in the form
of the quaternion  equation
\cite{Berezin-Kurochkin-Tolkachev-1989}
\begin{eqnarray}
\nabla \; [ ( \underline{B} - i \underline{E}) +(\underline{H} - i
\underline{D} )]+
  \stackrel{----------}
  {  \{ \nabla \;[ ( \underline{B} - i \underline{E}) -(\underline{H} - i \underline{D} )]   \} } ^{*} = 0 \; .
\label{1.4}
\end{eqnarray}

The Lorentz invariance of these equations is realized by the
following transforms\footnote{Quaternion $L$  corresponds to a
spinor matrix
 $B=k_{0} -i \vec{\sigma} \;\vec{k}\; \; \in SL(2.C)$  of  the  complex linear group
SL(2.C), spinor covering of the Lorentz group; restrictive
relation $L\;\bar{L}=1 $ is equivalent to separation special
linear group by imposing $\mbox{det}B = 1$.}
\begin{eqnarray}
\nabla ' =  L \nabla \bar{L}^{*}\; , \qquad \qquad  L \; \bar{L} =
k_{0}^{2} + \underline{k}^{2} = k_{0}^{2} + \vec{k}^{\;2} =  1  \;
, \nonumber
\\
(\underline{B}' - i \underline{E}') = L^{*} (\underline{B} - i
\underline{E}) \bar{L}^{*}\; , \qquad (\underline{H}' - i
\underline{D}') = L^{*} (\underline{H} - i \underline{D})
\bar{L}^{*}\; . \label{1.5}
\end{eqnarray}

The constitutive  relations in quaternionic form look
\begin{eqnarray}
(\underline{B} - i \underline{E}) = (\underline{H} - i
\underline{D} ) + [ ( \underline{H} + i \underline{D} ) \;
(\underline{\theta} + i \underline{\epsilon} ) ]_{S} \;
(\underline{H} - i \underline{D} )  + {1 \over 2} ( \underline{H}
+ i \underline{D} )^{2}_{S} \; (\underline{\theta} - i
\underline{\epsilon} ) \; , \nonumber
\\
(\underline{H} - i \underline{D}) = (\underline{B} - i
\underline{E} ) - [ ( \underline{B} + i \underline{E} ) \;
(\underline{\theta} + i \underline{\epsilon} ) ]_{S} \;
(\underline{B} - i \underline{E} )  - {1 \over 2} ( \underline{B}
+ i \underline{E} )^{2}_{S} \; (\underline{\theta} - i
\underline{\epsilon} ) \; .\nonumber
\\
\label{1.6}
\end{eqnarray}

Under the Lorentz group the quaternion $ \underline{\Phi}  =
(\underline{\theta} - i \underline{\epsilon} )$ transforms
according to the law
\begin{eqnarray}
\underline{\Phi}'  = L^{*}  \; \underline{\Phi}\;  \bar{L}^{*}\; .
\label{1.7}
\end{eqnarray}

We have arrived at a key relationship  determining  small Lorentz
group for  the noncommutativity object:
\begin{eqnarray}
L^{*}  \; \underline{\Phi}\;  \bar{L}^{*} = \underline{\Phi} \; ,
\qquad \mbox{or} \qquad L^{*}  \; \underline{\Phi}=
\underline{\Phi}\; L^{*} \;. \label{1.8}
\end{eqnarray}

\noindent it describes all inertial observers to which   effects
of non-commutativity will be seen exactly the same. It is
convenient to  introduce a new variable $\Phi^{*}=\varphi $, then
eq. (\ref{1.8})  reads
\begin{eqnarray}
(k_{0} +  \underline{k})   \; \underline{\varphi}=
\underline{\varphi}\; (k_{0} +  \underline{k}) \; . \label{1.9}
\end{eqnarray}

\noindent Evidently, that eq. (\ref{1.9}) is satisfied if and only
if quaternions $\underline{q}$ and $\underline{\varphi} $ are
proportional to each other:
\begin{eqnarray}
L_{\varphi} = (k_{0} + w \; \underline{\varphi}) \; , \qquad
\underline{\varphi} =
 (\underline{\theta} + i \underline{\epsilon} ) \; .
\label{1.9}
\end{eqnarray}

To proceed further in description of the subgroup (\ref{1.9}) we
should distinguish between two cases: $\underline{\varphi}^{2}
\neq 0$ and $\underline{\varphi}^{2} = 0$.

In the first case one can introduce new  parametrization in term
of a complex angle and unit quaternion:
\begin{eqnarray}
\underline{\varphi}^{2} \neq 0\;, \qquad k_{0} = \cos \chi \; ,
\qquad \underline{\varphi} =  \sqrt{\underline{\varphi}^{2} }  \;
{  \underline{\varphi}   \over  \sqrt{\underline{\varphi}^{2} }} =
\sin \chi \; \hat{\underline{\varphi}}\; , \qquad
\hat{\varphi}^{2} =  +1 \; ; \label{1.10}
\end{eqnarray}

\noindent so that the small Lorentz group  with  simple Abelian
multiplication law is given by
\begin{eqnarray}
L_{\varphi} = \cos \chi + \sin \chi \; \hat{\underline{\varphi}}\;
, \qquad \chi'' = \chi' + \chi \; .
\end{eqnarray}

Taking matrix realization for quaternion units, $e_{0}=I, e_{a} =
-i \sigma_{a}$, we get an explicit spinor form for  $L_{\varphi} $
\begin{eqnarray}
L_{\varphi}  = \cos \chi - i \; \sin \chi \;  \vec{\sigma} \; (
\vec{n} + i \vec{m}) \; , \qquad   \vec{n} + i \vec{m} = { (
\vec{\theta} + i \; \vec{\epsilon} ) \over \sqrt{ \vec{\theta}^{\;2}
- \vec{\epsilon}^{\;2} + 2i\; \vec{\theta} \; \vec{\epsilon} }} \; .
\label{1.10}
\end{eqnarray}

Immediately, particular examples when physical interpretation for
$\chi$-parameter is evident can be pointed out: Euclidean
rotations  and Lorentz boosts along vector $\vec{n}$:
 \begin{eqnarray}
\vec{n} \neq 0 \;, \; \vec{m}= 0\; , \;  \chi = \alpha + i \; 0\;
,\qquad L_{\varphi}  = \cos \alpha  - i \; \sin\alpha \;
\vec{\sigma} \;  \vec{n}  \;; \nonumber
\\
\vec{n} \neq 0 \;, \; \vec{m}= 0\; , \; \chi = 0 + i\; \beta \;
,\qquad \;\;\; \; L_{\varphi}  = \mbox{ch}\; \beta  +  \;
\mbox{sh} \; \beta \;  \vec{\sigma} \;  \vec{n}  \; .
\end{eqnarray}

\noindent It should be added, that the case of arbitrary
nonisotropic  $\theta^{\mu \nu}$ with the help of special Lorentz
transformation can be translated to a more simple form properties:
\begin{eqnarray}
\vec{n} + i \vec{m} \;, \qquad \vec{n}^{2} - \vec{m}^{2} = 1=
\mbox{inv}  \; , \;\;  \vec{n}\; \vec{m} = 0=\mbox{inv}  \qquad
\Longrightarrow \nonumber
\\
 \vec{n}'  + i \; 0 \;, \qquad  \vec{n}^{'2}= 1  =\mbox{inv} \; , \qquad
 L'_{\varphi}  = \cos \chi + i \; \sin \chi \;  \vec{\sigma}  \;  \vec{n}' \;  .
\nonumber
\end{eqnarray}

Thus, for arbitrary $\theta^{\mu \nu}$-tensor we arrive at the
corresponding small Lorentz group $SO(2)  \otimes SO(1,1)$,  just
this structure was previously described with   special choice of
non-commutativity matrix in \cite{Alvarez-Vazquez-2003},
\cite{Chaichian-Kulish-Nishijima-Tureanu-2008}.

In the second \underline{(isotropic)} case the normalization
condition  $L\; \bar{L}=1 $ gives
\begin{eqnarray}
\underline{\varphi}^{2} = 0\; ,  \qquad k_{0}^{2} + w^{2}  \;
\underline{\varphi}^{2} = 1 \;, \qquad k_{0} = \pm 1\; ; \nonumber
\end{eqnarray}

\noindent therefore now the small Lorentz group (see \ref{1.9}) )
is  specified by  relations
\begin{eqnarray}
L = \pm (1 + w\; \underline{\varphi} ) \; , \qquad
\underline{\varphi}^{2}= 0\; , \qquad  w'' = w' + w \; ,
\label{1.13}
\end{eqnarray}

\noindent where $w$ is any  complex  number. This is an Abelian
group of displacements in complex plane, $T_{2}$.

For   readers  preferring  the Maxwell theory in vector notation
let us give translation  to this language. Electromagnetic vectors
make up two complex 3-vector under complex orthogonal group
$SO(3.C)$:
\begin{eqnarray}
\vec{f} =   \vec{B}  -i\;  \vec{E}    \; , \qquad  \vec{h} =
\vec{H}   - i \; \vec{D}  \;  ,\qquad \vec{K} =  \vec{\epsilon} -
i \; \vec{\theta}  \; . \label{3.1a}
\end{eqnarray}

\noindent Complex orthogonal group may be defined as $2
\rightarrow 1$ mapping from  $SL(2.C)$ -- their elements are given
by
\begin{eqnarray}
SO(3.C)\; , \qquad O(k)  =\left | \begin{array}{lll}
 1 -2 (k_{2}^{2} + k_{3}^{2})   &   -2k_{0}k_{3} + 2k_{1}k_{2}    &   +2k_{0}k_{2} + 2k_{1}k_{3}  \\
 +2k_{0}k_{3} + 2k_{1}k_{2}     &  1 -2 (k_{3}^{2} + k_{1}^{2})   &   -2k_{0}k_{1} + 2k_{2}k_{3}   \\
 -2k_{0}k_{2} + 2k_{1}k_{3}     &   +2k_{0}k_{1} + 2k_{2}k_{3}    &  1 -2 (k_{1}^{2} + k_{2}^{2})
 \end{array} \right |
 \label{3.1b}
 \end{eqnarray}

\noindent governs their behavior under Lorentz group in accordance
with $ O(k) \;  \vec{f} = \vec{f}\; '\; , \; O(k) \;  \vec{h} =
\vec{h}' . $ One may note straightforwardly an identity
\begin{eqnarray}
O(k_{0}, \vec{k} ) \lambda \; \vec{k} =  \lambda \; \vec{k} \;
\label{3.1c}
\end{eqnarray}

\noindent which  is a base to explore the problem of small groups
for complex 3-vectors.

 \section{ On    discrete  dual symmetry }

Let us turn to Maxwell equation in quaternion form and rewrite
them as follows
\begin{eqnarray}
\nabla ( \underline{f} + \underline{h}) +
\stackrel{--------}{[\nabla ( \underline{f} -
\underline{h})]^{*}}= 0 \; , \label{2.1}
\end{eqnarray}

\noindent where $\underline{f} =  \underline{B} - i \underline{E}
\; , \; \underline{h} =  \underline{H} - i \underline{D} \;$ .
This equation is invariant under dual rotation:
\begin{eqnarray}
 \underline{f} + \underline{h}  = \underline{G}\; , \qquad  \underline{G}' = e^{i \chi} \; \underline{G} \;  ,
\nonumber
\\
 \underline{f} - \underline{h}  = \underline{R}\; , \qquad  \underline{R}' = e^{-i \chi} \; \underline{R} \;  .
\label{2.2}
\end{eqnarray}

\noindent We adhere to \cite{Abe} and take dual rotation for
$\underline{K} = \underline{\theta} - i \underline{\epsilon}$ in
the form $ \underline{K} ' = e^{i \chi} \; \underline{K} \;$. In
these variables, the constitutive relations read
\begin{eqnarray}
\underline{f}= \underline{h} + (\underline{h}^{*}
\underline{K}^{*})_{s} \underline{h}+ {1 \over 2} ( \underline{h}
^{*} \underline{h}^{*})_{s}\; K \; , \nonumber
\\
\underline{h}= \underline{f} - (\underline{f}^{*}
\underline{K}^{*})_{s} \underline{f}- {1 \over 2} ( \underline{f}
^{*} \underline{f}^{*})_{s}\; K \; . \label{2.3}
\end{eqnarray}

\noindent Summing and subtracting these two equations we get
\begin{eqnarray}
0 =  -{1 \over 2} ( \underline{R}^{*} \underline{K}^{*})_{s} G +
{1\over 2} (\underline{G}^{*} \underline{K}^{*})_{s} \underline{R}
- {1 \over 2} ( \underline{G}^{*} \underline{R}^{*})_{s}
\underline{K} - {1 \over 2} (\underline{R}^{*}
\underline{G}^{*})_{s} K \; , \nonumber
\\
2\underline{R} = {1 \over 2} ( \underline{R}^{*}
\underline{K}^{*})_{s} R  +  {1\over 2} (\underline{G}^{*}
\underline{K}^{*})_{s} \underline{G} + {1 \over 2} (
\underline{G}^{*} \underline{G}^{*})_{s} \underline{K} + {1 \over
2} (\underline{R}^{*} \underline{R}^{*})_{s} K \; . \label{2.4}
\end{eqnarray}

Requiring  invariance of these two relation with respect to dual
rotation
\begin{eqnarray}
\underline{G}= e^{-i\chi} \underline{G}' \; , \qquad
\underline{G}^{*} = e^{i\chi} \underline{G}^{'*} \; , \nonumber
\\
\underline{R}= e^{i\chi} \underline{R}' \; , \qquad
\underline{R}^{*} = e^{-i\chi} \underline{R}^{'*} \; , \nonumber
\\
\underline{K} = e^{-i\chi} \underline{K'} \; , \qquad
\underline{K}^{*} = e^{i\chi} \underline{K}^{'*} \; , \nonumber
\end{eqnarray}

\noindent we arrive at two equations: $ e^{i\chi} = e^{-3i\chi} \;
, \;  e^{-i\chi} = e^{+3i\chi} $
 with simple solution:
$ e^{i\chi} = 1,-1, +i, -i \; . $ Therefore, only discrete dual
transformation leaves invariant the nonlinear constitutive
equations,
 it corresponds to  $e^{i\chi}= \pm i $.
 Thus, the dual symmetry's status in noncommutative electrodynamics
differs  with that in  ordinary   linear Maxwell theory in
commutative space, this fact is  to be  interpreted in physical
terms.

%\section{Acknowledgement}

\vspace{5mm}

Author are grateful to Professor Kurochkin Ya.A. for discussion
and advice. This  work was  partially  supported  by the Fund for
Basic Research of Belarus, grant F08R-039.

\end{document}